\documentclass{aastex631} 
\pdfoutput=1

\usepackage{bm} 
\usepackage{amsmath} 
\usepackage{graphicx}
\usepackage{pgf}
\usepackage{layouts}
\usepackage{natbib}
\usepackage[caption=false]{subfig}

\DeclareMathOperator\arctanh{arctanh}

\begin{document}

\title{Gravelamps: Gravitational Wave Lensing Mass Profile Model Selection} 

\author{Mick Wright}
\affiliation{SUPA, School of Physics and Astronomy, University of Glasgow}

\author{Martin Hendry}
\affiliation{SUPA, School of Physics and Astronomy, University of Glasgow} 

\begin{abstract} 

	We present the package \textsc{Gravelamps} which is designed to analyse gravitationally-lensed gravitational wave signals in order to constrain the mass density profile of the lensing object. \textsc{Gravelamps} does this via parameter estimation using the framework of \textsc{bilby}, which enables estimation of both the lens and the source parameters. The package can be used to study both microlensing and macrolensing cases, where the lensing mass distribution is described by a point mass and extended mass density profile respectively. It allows the user to easily and freely switch between a full wave optics and approximate geometric optics description. The performance of \textsc{Gravelamps} is demonstrated via simulated analysis of both microlensing and macrolensing events, illustrating its capability for both parameter estimation and model selection in the wave optics and hybrid environments. To further demonstrate the utility of the package, the real gravitational-wave event GW170809 was analysed using \textsc{Gravelamps}; this event was found to yield no strong evidence supporting the lensing hypothesis, consistent with previously published results. 

\end{abstract} 

\section{Introduction} 

For most of the time that we have spent observing the universe, we have been able to do so using only the electromagnetic (EM) spectrum -- firstly via the visible window and then in the 20th century expanding to cover the full range from gamma rays to radio. However, with the dawn of gravitational-wave (GW) astronomy and the first detections of signals from the mergers of compact binary systems made by LIGO and VIRGO, we have now opened an entirely new gravitational-wave window on the universe. The detection of gravitational waves was the culmination of a century of research, from the first theoretical description of the phenomenon \citep{1916SPAW.......688E} in 1916 up to the observation of GW150914 \citep{PhysRevLett.116.061102}, and onwards -- with now approximately 100 confirmed detections \citep{Abbott_2019} \citep{theligoscientificcollaboration2021gwtc21} \citep{Abbott_2021}, and the prospect of many hundreds more as the extended 2nd-generation ground-based detector network reaches its design sensitivity later this decade.

In addition to the development of the ground-based detectors themselves, in recent years a great deal of work has gone into creating computational tools capable of generating predicted GW waveforms. These are compared to interferometer data, in order to detect sources and infer their parameters -- as discussed in more detail in \cite{Abbott_2020}. Two extensive toolkits of particular importance are the LIGO Algorithm Library, or \textsc{LAL} \citep{lalsuite},  and \textsc{Bilby} \citep{2019ApJS..241...27A}: the former allows the straightforward generation of many different types of waveform, as just one of its many functionalities, whilst the latter presents an easy interface with which to perform the aforementioned parameter estimation using a variety of nested sampling \citep{2004AIPC..735..395S} \citep{10.1214/06-BA127} algorithms.

A recent focus of attention for the nascent field of GW astronomy has been the gravitational lensing of gravitational waves \citep{theligoscientificcollaboration2021search} \citep{broadhurst2019twin} \citep{refId0} \citep{Janquart_2021} \citep{seo2021strong}. Just as light passing by a massive compact object is deflected by the warping of space-time, so too gravitational wave signals are affected by the distribution of intervening matter as they propagate towards us. Electromagnetic observations of gravitational lensing have played an important role in the development of general relativity, with the deflection of light being one of its four major tests \citep{1923MmRAS..62A...1D}. In a similar way, it is hoped that future observations of the lensing of gravitational waves may prove to be not just an important further test of general relativity and other theories of gravity, but also a powerful diagnostic probe of the nature of dark matter \citep{mishra2021gravitational} \citep{urrutia2021lensing}.

Whilst dark matter may not be probed directly using light, EM observations of gravitational lensing have been used to reconstruct the distribution of dark matter in the lensing object \citep{10.1093/mnras/283.3.837} \citep{amorisco2021halo} and a number of authors have recognised that GW gravitational lensing could also provide a valuable adjunct to these already extant investigations \citep{wang2021lensing}. Already, authors have investigated how the mass density profile will affect the properties of lensed GW signals \citep{PhysRevD.90.062003} \citep{Takahashi_2003}, and sought to demonstrate that these properties may be used to characterise the lensed events -- thus constraining the lens model parameters \citep{Herrera_Mart_n_2019} \citep{mishra2021gravitational}. By considering different lens models, which correspond to differing kinds of lensing objects such as isolated point masses, dark matter halos, etc, analysis of the lens model therefore also provides insight into the nature of the lensing object. 

Extending the above work, presented here is the package \textsc{Gravelamps}, designed to analyse lensed gravitational-wave data using arbitrary lens models. Built upon the \textsc{LAL} and \textsc{Bilby} frameworks, \textsc{Gravelamps} performs parameter estimation on GW events that have already been confidently identified as lensed, based on the application of such pipelines as e.g. \textsc{Golum} \citep{Janquart_2021} or \textsc{hanabi} \citep{lo2021bayesian}. Specifically, \textsc{Gravelamps} can analyse a range of different mass density profile models, yielding estimates for the joint posterior distribution of the lens and source parameters for each model. The code also allows comparison between different models through calculation of their relative Bayes Factor, which quantifies how probable one model is compared to another, thus providing a means to quantitiatively evaluate which mass profile for the lens is most likely. 

In the following sections, the reader will be guided through the theory behind \textsc{Gravelamps} and its usage, performance, design sensitivities, as well as results from it and how it may be extended. Section 2 will cover the theory of gravitational lensing in general, Section 3 will cover the specific profiles that are presently integrated into \textsc{Gravelamps}, Section 4 will develop upon this to describe the calculation of the amplification factor, Section 5 will then go into more detail about the code itself -- giving more detailed explanations of how it was constructed and how it can be used. Section 6 will then present some results from usage of \textsc{Gravelamps} in a variety of possible scenarios. Finally, Section 7 will conclude and give some of the aforementioned detail on possible future extensions, as well as integration of the code into existing analysis pipelines. 

\section{Gravitational Wave Lensing} 

For an overview of gravitational lensing theory, the reader is referred to works such as \citep{mollerach2002gravitational} or \citep{schneider2013gravitational}. Here we give a very brief introduction to the gravitational lensing of GWs.  In this case the relationship between a lensed signal, $h^{L}\left(f\right)$, and an unlensed signal, $h\left(f\right)$, takes the form \citep{Takahashi_2003}: 

\begin{equation}
	h^{L}\left(f\right) = h\left(f\right) \times F\left(f\right)
	\label{lensedunlensed}
\end{equation} 

where the quantity $F\left(f\right)$ is called the \textit{amplification factor} and is the ratio between the wave amplitudes of the observed lensed $\phi^{L}_{obs}\left(w, \bm{\eta}\right)$ and unlensed, $\phi_{obs}\left(w, \bm{\eta}\right)$ signals: 

\begin{equation}
	F\left(w, \bm{\eta}\right) = \frac{\phi^{L}_{obs}\left(w, \bm{\eta}\right)}{\phi_{obs}\left(w, \bm{\eta}\right)}
	\label{ampfacdef} 
\end{equation} 

where $w$ is the dimensionless frequency, and $\bm{\eta}$ as defined above is the position of the source. The simplicity of the relation between the lensed and unlensed signal is owed to the fact that with the exception of the frequency, the source parameters defining the unlensed signal and the lens parameters defining the amplification factor are uncorrelated. What does determine the amplification factor, however, is the mass density profile of the lens object. This will be explored in more detail in the next section. 

\section{Lensing Profiles and the Amplification Factor}

As was noted in the introduction, lensing objects can be a number of differing things: from isolated point masses to extended objects like dark matter halos. One of the ways in which these differences manifest is in the distribution of the mass of the object given by $\rho(\bm{x})$. These density profiles result in different lensing signatures \citep{Takahashi_2003}, which in principle means that the lensing signature can provide information on the density profile. Thus, in principle, it is possible to determine whether a model profile is consistent with observed lensed data thus meaning that such an analysis can place important constraints on the nature of the lensing object. This determination of the model profile is that \textsc{Gravelamps} seeks to make. 

Whilst the density profile is different for each lens model and this ultimately yields different forms of the amplification factor, the underlying mathematics is the same. The astrophysical distances involved allow the use of the so called thin-lens approximation to be applied. This means that instead of requiring the full three dimensional mass density profile, $\rho(\bm{x})$, the surface mass density, $\Sigma(\bm{\xi})$ may be considered instead. This is the projection of the three-dimensional density onto a two-dimensional plane perpendicular to the line-of-sight to the source and at the distance of the centre of mass of the lensing object, i.e \citep{mollerach2002gravitational} 

\begin{equation}
	\Sigma(\bm{\xi}) = \int \rho \left(\bm{\xi}, z\right) dz. 
	\label{surfmassden}
\end{equation} 

Should the extended mass distribution be spherically symmetric, some further simplification may be made. In this case, the surface mass density depends only upon the modulus of the impact parameter, $\xi = \left|\bm{\xi}\right|$. 

Using an arbitrary normalisation length, $\xi_0$, one may construct dimensionless quantities from the impact parameter and the source position \citep{schneider2013gravitational}:

\begin{equation}
		\bm{x} = \frac{\bm{\xi}}{\xi_0} \quad , \quad  \bm{y} = \frac{D_{OL}}{\xi_0D_{OS}}\bm{\eta},
	\label{dimquant}
\end{equation}

with $D_{OL}$ and $D_{OS}$ being the angular distances between the observer and the lens and source respectively. Whilst the value of the amplification factor differs between the lensing profiles, it can be caculated from a general expression using the above dimensionless quantities \citep{Takahashi_2003} \citep{schneider2013gravitational}:

\begin{equation}
	F\left(w, y\right) = \frac{w}{2\pi i} \int d^2x \exp\left[iwT\left(x,y\right)\right],
	\label{ampfacgeneral}
\end{equation} 

where $T\left(x,y\right)$ is the dimensionless time delay. The simplest form of this general expression is for the case of axially-symmetric lensing object, where it is given by \citep{Takahashi_2003}:

\begin{equation}
	\begin{aligned}
	F\left(w,y\right) = -iwe^{iwy^2/2} \int_{0}^{\infty} dx x J_0 \left(wxy\right) \\  
	\times \exp \left[ iw\left(\frac{1}{2}x^2 - \psi(x) + \phi_{m}(y)\right)\right],
	\end{aligned}
	\label{ampfacaxisym}
\end{equation}

where $\psi(x)$ is the lensing potential, $\phi_m(y)$ is the phase for the minimum time delay - chosen such that the minimum time delay is zero - and $w$ is the dimensionless frequency defined by \citep{Takahashi_2003}:

\begin{equation}
	w = \frac{D_{OS}}{D_{LS}D_{OL}} \xi_0^2 \left( 1 + z_{l} \right) \omega.
	\label{dimfreq}
\end{equation} 

In the case where the dimensionless frequency is very high (i.e. $w >> 1$), the geometric optics approximation may be made. In this case the stationary points of the dimensionless time delay are the only contributors to the amplification factor integral, yielding a simpler expression for the amplification factor \citep{10.1143/PTPS.133.137}: 

\begin{equation}
	F_{geo} \left(w,y\right) = \sum_{j} \left|\mu_j\right|^{1/2} \exp \left[iwT_j -i\pi n_j\right], 
	\label{ampfacgeometric}
\end{equation} 

where $\mu_j$ denotes the magnification of the $j^th$ image, $T_j = T(\bm{x}_j, \bm{y})$, and $n_j = 0, 1/2, 1$ when $\bm{x}_j$ is a minimum, saddle point, or maximum of $T(\bm{x},\bm{y})$ respectively. 

\subsection{Point Mass}

The simplest possible lens case is that of a point mass. It is applicable in the cases of compact objects such as individual stars or black holes. It can also be used for extended objects, by means of the Birkhoff Theorem, in the case where the lens size is much smaller than the Einstein radius. It is one of the only cases in which the full wave optics calculation of the amplification factor, i.e. Equation \ref{ampfacaxisym}, can be carried out analytically. Here the normalisation constant $\xi_{0}$ introduced above corresponds to the Einstein radius and is given by:

\begin{equation}
	\xi_0 = \left( \frac{4M_LD_{OL}D_{LS}}{D_{OS}}\right)^{1/2}.
	\label{pointmassnorm}
\end{equation}

This leads to the dimensionless frequency being given by $w = 4M_{lz}\omega$ where $M_{lz}$ is the redshifted lens mass. This ultimately leads to the amplification factor in the full wave optics case being given by \citep{Takahashi_2003}:

\begin{equation}
	\begin{aligned}
	F(w,y) = \exp \left[\frac{\pi w}{4} + i\frac{w}{2} \left\{ \ln \left( \frac{w}{2} - 2\phi_m(y) \right) \right\} \right] \\
	\times \Gamma \left(1 - \frac{i}{2}w \right) {}_1F_1 \left(\frac{i}{2}w;1;\frac{i}{2}wy^2\right),
	\end{aligned}
	\label{pointlens_ampfac}
\end{equation} 

where ${}_1F_1$ is the confluent hypergeometric function of the first kind, and $\phi_m(y)$ is given by $(x_m - y)^2/2 - \ln x_m$ where $x_m = (y + \sqrt{y^2 + 4})/2$. In the geometric optics approximation case, the amplification factor is given by \citep{Takahashi_2003}

\begin{equation}
	F_{geo}(w,y) = \left| \mu_+ \right|^{1/2} - i \left| \mu_- \right|^{1/2} e^{iw\Delta T},
	\label{pointlens_geomfac}
\end{equation} 

where the magnifications are given by $\mu_{\pm} = 1/2 \pm (y^2 + 2)/(2y\sqrt{y^2 + 4})$ and the time delay between the two images is given by $\Delta T = y\sqrt{y^2 + 4}/2 + \ln((\sqrt{y^2 + 4} + y)/(\sqrt{y^2 + 4} -y)$. 

\subsection{Singular Isothermal Sphere}

The Singular Isothermal Sphere, or SIS, profile is the simplest and most widely used profile that has been designed to model the behaviour most commonly observed for galaxies - i.e. a flat rotation curve. This behaviour as a result of modelling the galaxy as a large, extended object containing luminous matter embedded within a dark matter halo. Whilst the key strength of the SIS profile is the ability to replicate this flat rotation curve \citep{2007ApJ...667..176G}, it does suffer from a weakness in the form of a central singularity which is non-physical \citep{Burkert_1995} \citep{2021}. The full SIS density profile is given by \citep{1987gady.book.....B}:

\begin{equation}
	\rho_{SIS}(r) = \frac{\sigma^2}{2\pi Gr^2},
	\label{sis}
\end{equation}

where $\sigma$ is the one-dimensional velocity distribution of stars around the galaxy being modelled. The normalisation chosen for the SIS profile, similarly to the point mass case, is the Einstein Radius, which in the SIS case with velocity disperison, $v$, is given by \citep{Takahashi_2003}:

\begin{equation}
	\xi_0 = 4\pi v^2 \frac{D_{OL}D_{LS}}{D_{OS}}.
	\label{sisnorm}
\end{equation}

The redshifted lens mass in the SIS profile is given by \citep{Takahashi_2003}:

\begin{equation}
	M_{lz} = 4\pi v^4 \left[ \left(1+z_l\right) \frac{D_{OL}D_{LS}}{D_{OS}} \right], 
	\label{sismlz}
\end{equation}

which results in a consistent definition of the dimensionless frequency, $w$ as with the point mass. 

The amplification factor may be calculated from Equation \ref{ampfacaxisym} by identifying that in the SIS case $\psi(x) = x$ and $\phi_m(y) = y + 1/2$. Expanding the second term in the exponential into an infinite sum yields:

\begin{equation}
	\begin{aligned}
	F(w,y) = -iwe^{\frac{i}{2}wy^2 + iw\phi_m(y)} \\
	\times \sum_{n=0}^\infty \frac{(-iw)^n}{n!} \int_0^\infty x^{1+n} J_0(wxy) e^{\frac{1}{2} iwx}
	\end{aligned}
	.
	\label{sismidstage}
\end{equation}

This equation is solvable. Using the identity that $e^z {}_1F_1(a,b;-z) = {}_1F_1(a,b;z)$ \citep{1965hmfw.book.....A}, this ultimately yields \citep{2006}: 

\begin{equation}
	\begin{aligned}
	F(w,y) = e^{\frac{i}{2}w(y^2 + 2\phi_m(y))}\sum_{n=0}^\infty \frac{\Gamma \left(1 + \frac{n}{2}\right)}{n!} \\
	\times \left( 2we^{i\frac{3\pi}{2}}\right)^{n/2} {}_1F_1 \left( 1 + \frac{n}{2}, 1; -\frac{i}{2}wy^2 \right) 
	\end{aligned}
	,
	\label{sis_amp_fac}
\end{equation}

In the geometric optics case, the source position determines the number of images that will be formed. If $y < 1$ double images occur, where if $y \geq 1$ only a single image is formed meaning that the amplification factor is split between these cases, being given by \citep{Takahashi_2003}:

\begin{equation}
	F_{geo} (w,y) =
	\begin{cases}
		\left|\mu_+\right|^{1/2} - i\left|\mu_-\right|^{1/2}e^{iw\Delta T} & \text{ for } y < 1 \\
		\left|\mu_+\right|^{1/2} & \text{ for } y \geq 1
	\end{cases},
	\label{sislens_geomfac}
\end{equation}

where in this case the magnficiations are given by $\mu_{\pm} = \pm 1 + 1/y$ and the time delays are given by $\Delta T = 2y$. 
\subsection{Navarro, Frenk, White} 

In the standard Concordance Model of Cosmology \cite{2020} \cite{2001} the dark matter is assumed to be cold and collisionless. Consistent with these properties, the Navarro, Frenk, and White, or NFW, profile is widely used to model the density profile of cold dark matter (CDM) halos, as described by a singular `universal' scaling function. Whilst allowing for more complexity than the Singular Isothermal Sphere profile, the NFW profile suffers from the same central singularity, or cusp. The density profile is given by \citep{1997ApJ...490..493N}:

\begin{equation}
	\rho_{NFW}(r) = \frac{\rho_{s}}{\frac{r}{r_s} \left(1 + \frac{r}{r_s}\right)^2},
	\label{nfw}
\end{equation}

where $\rho_s$ is the central density and $r_s$ is the characteristic scale of the profile. This characteristic scale is also chosen as the normalisation constant $\xi_0$ for this profile. 

Owing to the increasing complexity of the profile, the form of the lensing potential is more complicated being given by \citep{1996A&A...313..697B}:

\begin{equation}
	\psi(x) = \frac{\kappa_s}{2}
	\begin{cases}
		\left[ (\ln \frac{x}{2})^2 - (\arctanh \sqrt{1-x^2})^2 \right] & \text{ for } x \leq 1 \\
		\left[ (\ln \frac{x}{2})^2 - (\arctan \sqrt{x^2 -1})^2 \right] & \text{ for } x \geq 1
	\end{cases}.
	\label{nfwlenspot}
\end{equation}

Here, $\kappa_s$ is the dimensionless surface density which is given by $16\pi \rho_s (D_{OL}D_{LS}/D_{OS})r_s$. The time delay constant also becomes more complicated as the lens equation no longer has an analytical solution meaning it must be calculated numerically, meaning the amplification factor too must be calculated numerically from Equation \ref{ampfacaxisym} in the wave optics case, or Equation \ref{ampfacgeometric} in the geometric optics case. In the latter case, the number of images formed is dependent upon the value of $y$ relative to a critical value -- with three images being formed where $\left|y\right| < y_{cr}$, and only one in the case where $\left| y \right| > y_{cr}$. In the case of the source position being precisely at the critical value, the magnification becomes infinite. 

\section{Gravelamps Overview}

\textsc{Gravelamps} is an analysis pipeline designed to study arbitrary lensed gravitational-wave data and constrain properties of both the source and mass density profile of the lens . The pipeline estimates the Bayesian evidence for the user specified lens model. In so doing it also allows given that chosen model inference of both the source and lens parameters. By performing this analysis for multiple lens models, comparison may be made between these evidences to determine the most likely model for the data. The application of \textsc{Gravelamps} to some example scenarios, to illustrate its usefulness for identifying and characterising lensed gravitational-wave signals, will be presented in the following sections. 

\subsection{Languages and Toolkits Used}

For the higher level parts -- in terms of interacting with the user and performing the parameter estimation, \textsc{Gravelamps} is coded in \textsc{python}. This allows for flexibility and ease of reading as well as access to a number of useful modules. In use in \textsc{Gravelamps} are the extensively used \textsc{numpy} \citep{harris2020array}, \textsc{scipy} \citep{2020SciPy-NMeth}, and \textsc{astropy} \citep{2018AJ....156..123T} modules. Most importantly, the use of \textsc{python} gives access to the \textsc{bilby} package \citep{2019ApJS..241...27A} thus ensuring the use of both reliable parameter estimation methods and  native and easily extensible support for LAL waveforms. This creates a more cohesive experience for a user wishing to modify the program beyond the level of the options already available to them. The use of \textsc{bilby} also allows the pipeline to be scaled straightforwardly for usage on clusterised machines, through \textsc{bilby\_pipe} \citep{Romero_Shaw_2020}.  In particular, there is little difference between a run designed to be implemented on a local machine and one designed for a clusterised machine, e.g. to be submitted through the scheduler \textsc{HTCondor}, as is used widely in the LSC clusters.  

For the lensing data generation, as can be seen in the previous section, the amplification factor calculation is not a simple task as it requires the calculation of the integral of a highly oscillatory function. Due to the intensive nature of the calculations, these were coded in C++ to benefit from the additional speed of a compiled language -- as well as granting access to the C arbitrary precision library \textsc{arb} \citep{Johansson2017arb}. In addition to being able to perform the necessary calculations with appropriate speed and accuracy, this approach had the benefit of allowing users to decide for themselves how much they wished to trade said speed and accuracy, as well as allowing the full wave optics to be calculated to as high a dimensionless frequency value as possible -- well past the point at which the geometric optics approximation can successfully take over -- without causing disruption to the amplification factor that is calculated.

In addition to those toolkits used directly in \textsc{Gravelamps}, there are other pieces of software which are critical to the functioning of the toolkits themselves -- with the \textsc{bilby} package additionally depending upon such packages as \textsc{matplotlib} \citep{Hunter:2007} and \textsc{corner} \citep{corner} for plotting, and \textsc{gwpy} \citep{gwpy} and \textsc{lalsuite} \citep{lalsuite} for gravitational wave data analysis. Further, through \textsc{bilby}, \textsc{Gravelamps} allows the use of a great many nested samplers: chosen for the work presented here was \textsc{dynesty} \citep{2020MNRAS.493.3132S}. 

\subsection{Design Intentions}

As with almost all software packages, there were certain philosophies which underpinned how the software was constructed, in addition to the obvious requirements of functionality and speed. In the case of \textsc{Gravelamps}, the following were the most important considerations: 

\begin{enumerate}
	\item{\textit{Openness}: With such complicated integrals being necessary to perform the calculations to generate the amplification factor, it is tempting to consider platforms such as \textsc{Mathematica} which are designed to tackle extremely complex mathematics easily and with speed. However, a problem with such platforms is their proprietary nature. \textsc{Gravelamps} is designed to be open for use in as many applications as possible without any licensing conflicts; as such the toolkits used were chosen to reflect this, as well as the license chosen for \textsc{Gravelamps} itself -- with the source code for all of these being freely available \citep{gravelamps_repo}}.

	\item{\textit{Simplicity}: In the spirit of the openness with which \textsc{Gravelamps} was designed, it was also designed to be as understandable to the user as possible. In particular, the code was designed with ease of readability in mind, and so that interactions with the software by the user should be as simple as possible. As such, there is little difference between a run on a local machine and that on a cluster -- requiring only manipulation of the configuration INI file.} 

	\item{\textit{Extensibility}: Whilst at this time, only the three density profiles discussed in detail in this paper are fully coded, \textsc{Gravelamps} as a platform is designed to be easily extensible -- for example by including other density profiles or by considering multiple lenses.}
\end{enumerate}

\subsection{Structure of an Analysis Run}

Having been designed to be relatively simple, \textsc{Gravelamps} does not require that the user in any way modify the backend, in order to carry out an analysis run; instead the user may select all necessary settings through means of a simple INI file -- contained within which there are options governing:

\begin{itemize}
	\item{The output information}
	\item{Lensing settings - such as the start, end and, number of points for the amplification factor grid if it being used as well as those settings pertaining to arithmetic precision}
	\item{\textsc{HTCondor} settings for clusterised runs}
	\item{\textsc{bilby\_pipe} settings for clusterised runs}
	\item{Common analysis options such as the waveform model adopted (including whether to use direct or interpolated calculation of the amplification factor), trigger time chosen, etc}
	\item{Whether or not to perform an unlensed analysis, and if so, the associated settings to be used}
	\item{Event/Injection parameters}
	\item{Sampler settings} 
\end{itemize}

Here the user is given as much control over the run as possible without the requirement of delving into the code and modifying it directly for each analysis - although the latter is, of course, possible too. With a completed INI, the user need only call one of the preconstructed analysis functions, \texttt{gravelamps\_local\_inference} or \texttt{gravelamps\_pipe\_inference}, depending on whether they wish the run performed locally, or they wish submit files for the \textsc{HTCondor} Scheduler prepared. 

Should users be implementing a clusterised version of the code, they will receive a DAG file that will call each of the necessary submit files in order. If running locally, the steps will be carried out directly. From the point at which the analysis run begins, the procedure is as follows:

\begin{enumerate}
	\item{Over the specified ranges of the dimensionless frequency and source position, generate a grid of values of the amplification factor for the chosen lens model}
	\item{Generate an interpolating function to evaluate the amplification factor for any given dimensionless frequency and source position within the specified ranges}
	\item{Generate injection data/Fetch event data -- i.e. prepare the data that will be analysed}
	\item{Run parameter estimation on the data using the chosen lensed waveform model}
\end{enumerate}

In the case where the user is operating solely in the geometric optics regime, to assist with computational efficiency they may also choose to have the amplification factor calulated directly in the waveform, bypassing the need for steps 1 and 2. 

The procedure outlined above results in acquiring both estimates of the lens and source parameters for the given model, and also an estimate of the evidence for that model. This can then be compared with the evidences for other models to give a quantitative indication of the preferred model. This latter step may include, if the user wishes, an estimate of the evidence for an unlensed model.  In this unlensed case, however, \textsc{Gravelamps} should not be considered as providing a true Bayes factor for the relative probability of the lensed and unlensed hypotheses since no explicit evaluation of any selection effects relevant to the lensed-to-unlensed posterior odds ratio is carried out, based on e.g. a numerical injection study that models the probability of `false positives' -- i.e. pairs of candidate gravitational-wave events with parameters that could be consistent with them being strongly-lensed multiple images of a single source \citep{theligoscientificcollaboration2021search}. In its current form, \textsc{Gravelamps} conducts lens and source model parameter estimation on only a single such image.  In the future we plan to explore its extension to the framework appropriate for analysis of multiple images, but in the remainder of this work we consider only the single image case.

\subsection{Package Structure} 

Following the standard \textsc{python} package layout of submodules with specific purposes in mind, \textsc{Gravelamps} is split into \texttt{gravelamps.inference} and \texttt{gravelamps.lensing} covering the performance of the inference runs and the lensed waveform generation respectively. 

\subsubsection{\texttt{gravelamps.inference}}

Contained within \texttt{gravelamps.inference} are the individual scripts that form the programs by which the user interacts with \textsc{Gravelamps}, i.e. \texttt{gravelamps\_local\_inference} and \texttt{gravelamps\_pipe\_inference}. In addition to the programs themselves, contained with \texttt{gravelamps.inference} are all of the functions which are concerned with handling the user's configurations as well as the generation of the \textsc{HTCondor} submit files for the clusterised code. These are contained within the \texttt{helpers} and \texttt{file\_generators} parts of the submodule. 

\subsubsection{\texttt{gravelamps.lensing}}

\texttt{gravelamps.lensing} contains within it those parts of the source code pertaining to the construction of lensed waveforms. It is itself split between a \textsc{C++} and a \textsc{python} component. It also in a similar manner to the \texttt{gravelamps\_local\_inference} or \texttt{gravelamps\_pipe\_infernece} contains a pair of programs -- \texttt{gravelamps\_generate\_lens\_local} and \texttt{gravelamps\_generate\_lens\_pipe} that will generate lensing data without then constructing a lensed waveform and performing analysis on it. 

The \textsc{C++} parts of this submodule contains the source codes for the functions that calculate the amplification factor values, as well as the program that generates amplification factor data from which an interpolator may be constucted. These programs are compiled upon the installation of \texttt{Gravelamps}. Each of these programs functions in a similar maner: taking in as inputs dimensionless frequency and source position information contained within unique files - together with user defined settings, such as the dimensionless frequency value at which to begin using the geometric optics approximation, or the arithmetic precision of the wave optics calculations. Eaah program then returns two files containing two unique files containing the amplification factors' real and imaginary components, which can then be used to generate a complex interpolator for the amplification factor function. 

The \textsc{python} parts of the submodule concern both the utility functions that pertain to the construction of the dimensionless frequency and source position data used by the above. Most importantly, it also contains those functions that construct the lensed waveforms over both the wave and geometric optics regimes. This is done by firstly either using the data generated by the \textsc{C++} programs to generate an interpolating function for the amplification factor, or should the user wish to work exclusively in geometric optics they may also instruct \texttt{gravelamps.lensing} to instead, by means of the \textsc{ctypes} module contained with \textsc{python}, directly use the calculation function from the \textsc{C++} part. Once the amplification factor calculation function is established, it will then generate a base LAL waveform and using the amplification factor lens this waveform. Finally, within the \textsc{python} section, are those functions which perform physical calculations such as unit conversion, etc, whereas those calculations in \texttt{gravelamps.inference} are more statistical in nature. 

Contained within the \textsc{python} parts of \texttt{gravelamps.lensing} are the utility functions that pertain to the construction of the lensed waveform directly, such as generating the dimensionless frequency and source position data over which interpolation will take place, as well as taking the base unlensed LAL waveforms and turning them into the corresponding lensed versions. In addition, \texttt{gravelamps.lensing} contains those functions which perform physical calculations such as unit conversion, etc, whereas those calculations in \texttt{gravelamps.inference} are more statistical in nature.

The ability to implement both the wave optics and geometric optics calculations is particularly useful, in view of the complexity of the calculations required for the full wave optics analysis. Whilst these can be run for higher values of the dimensionless frequency, the calculation requires ever increasing precision and greater numerical limits on the integrations and summations; this leads to an increasing slow down due to the necessity of performing more calculations to reach these higher thresholds. However, \textsc{Gravelamps} has been written in a sufficiently flexible state to leave this choice to the user rather than making it for them.

\section{Example Uses of Gravelamps}

One of the major focuses of \textsc{Gravelamps} is to be as versatile in terms of how it can be used to study lensing of gravitational wave signals. As such, the lensing data can be generated by themselves, i.e. simulated data can be analysed, as well as real data. The following will cover examples of each of these use cases to demonstrate how an analysis might be made.  

\subsection{Generation of Amplification Factor Data} 

\begin{figure*}[ht!]
	\includegraphics[width=\textwidth]{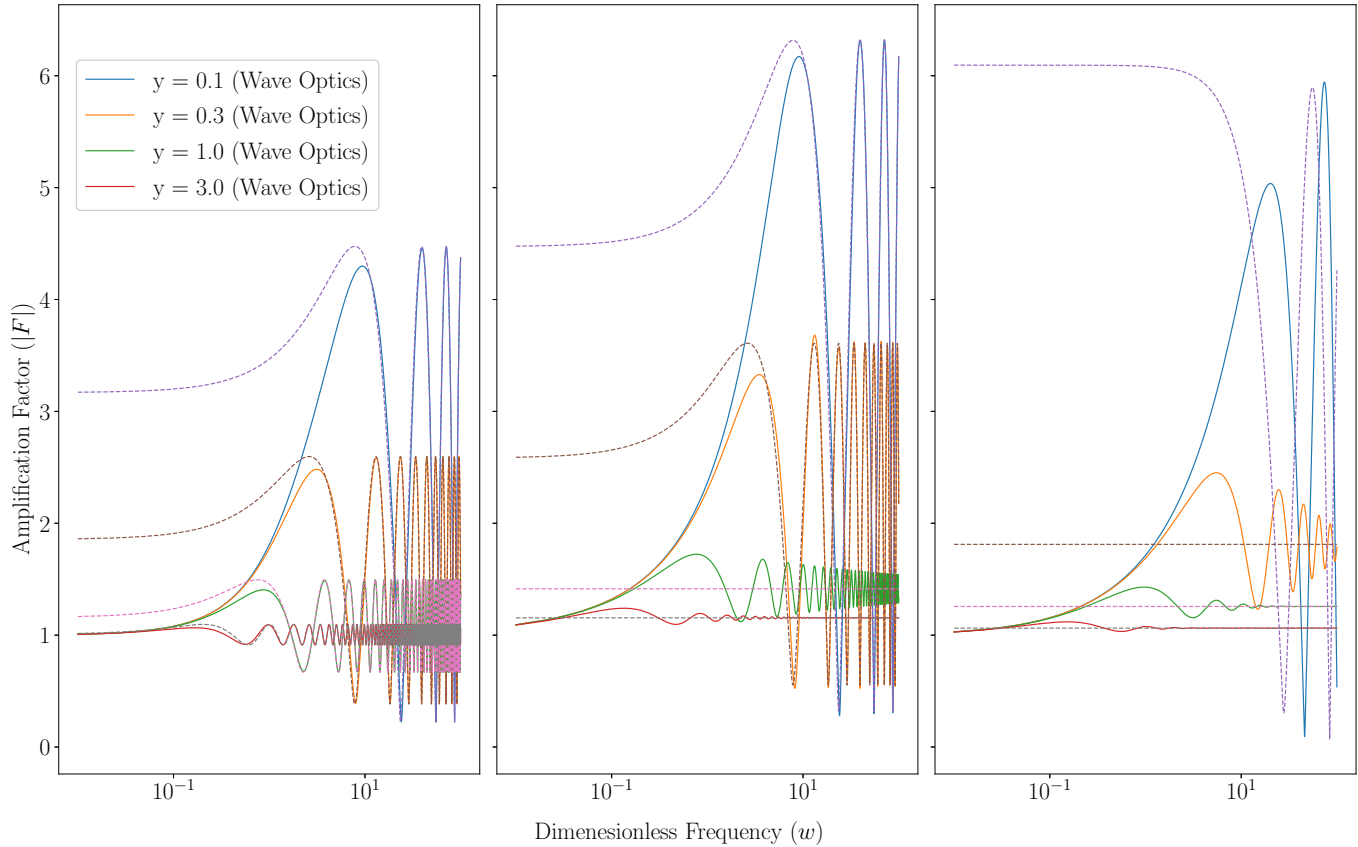}
	\caption{Absolute values of the amplification factor as calculated by Gravelamps for (from left-to-right): the point lens model, the SIS model, and the NFW model. The solid lines are those values calculated for the wave optics approach, with the dashed lines for those calculated using the geometric optics approximation.}
	\label{amp_fac_plot}
\end{figure*}

As was discussed in the previous sections, the generation of a lensed waveform from an unlensed one is simply a matter of multiplying the strains of the unlensed waveform by the amplification factor function corresponding to the lensing model of interest. With the base, unlensed, waveforms being implemented by \textsc{LAL}, \textsc{Gravelamps} generates the amplification factor to then lensed these waveforms. As was also previously noted, there are two functions that can caclulate the amplification factor: the full wave optics approach or the geometric optics approximation. Both of these have benefits and drawbacks, and both are implemented within \textsc{Gravelamps}. 

Figure \ref{amp_fac_plot} shows the results of using \textsc{Gravelamps} lens generation codes for both the wave and geometric optics cases, up to a dimensionless frequency of $w = 100$ over a range of source position values, for each of the Point Mass, SIS, and NFW lens models to illustrate the differences and applicability of both calculation styles. 

In the case of full wave optics, the calculations are obviously more complete. However, they are significantly more computationally intensive, requiring the use of arbitrary precision calculations as well as the use of numerical integration in the most complex case. Even the SIS - the simplest of the extended object models requires the approximation of an infinite sum to a finite degree. This makes the generation of the wave optics amplificaiton factor slower -- increasingly so with increasing dimensionless frequency where in order to retain numerical stability of the result, greater limits need to be placed on the integrations as well as increasing precision.

In the case of the geomtric optics approximation, the calculations are much simpler -- resulting in sufficient speed that they may be used directly in the waveform generation as opposed to the interpolator approach used in the wave or mixed approaches. However, strictly this approximation is only applicable at higher dimensionless frequency values as at lower dimensionless frequency values they do not replicate the behaviours of the full wave optics function -- noting that in the SIS and NFW cases the damped oscillatory nature of the higher source position value curves are replaced with a signle value that approximates the average of the oscillation. The increase in speed however, does warrant the use of geometric optics in higher dimensionless frequency regimes. 

\textsc{Gravelamps} was designed to be a versatile system however, so the user is given as much control over these calculations as possible. The user from the INI file used to generate the amplification factor data is able to specify firstly whether to use the direct or interpolation methods. The former may be used for geometric optics alone, where the latter may be used for wave optics alone, or a mix of wave and geometric optics, (or indeed also for geometric optics alone). In the case of using the interpolator method, the INI file used to generate the amplification factor data is able to specify:

\begin{itemize}
	\item{The minimum and maximum dimensionless frequency and source position}
	\item{The number of dimensionless frequency and source position to be generated for the interpolator}
	\item{The precision used in the arbitrary precision calcualtions (for wave optics only)}
	\item{The upper limit of the summation and integration used in the SIS and NFW models respectively}
\end{itemize}

These choices are given to the user to allow them to decide for themselves how much speed/accuracy trading that they wish to adopt. This also allows for a hybrid approach, in particular preventing the user from being forced to use e.g. geometric optics in all cases; if the user has the computational resources and time available they are able to continue using the wave optics approach to as high a dimensionless frequency as they wish. It should be noted, however, that the higher dimensionless frequency wave optics calculations become ever more computationally expensive in order to retain numerical stability. This increase becomes particularly noticable between $w=100$ and $w=1000$. It is therefore recommended that should the user need to use wave optics above these values that they do so only when they are planning to use a single lens generation for multiple parameter estimation runs where the justification of such computational expense may be split between the runs.

The flexibility of \textsc{Gravelamps} extends to other features. The user is able to specify any terminal-callable process in the INI and also what arguments are needed, allowing the user the ability to generate new models easilyt, even without adding them to \textsc{Gravelamps} itself directly (although users are encouraged to incorporate any such new models into the codebase should they wish to do so). If the user chooses to use non-implemented models, they are able to use the \texttt{gravelamps\_generate\_lens} programs to generate amplification factor data. Placing the locations of the resultant files into the INI's optional input then allows these data to be accessed in inference runs.

\subsection{Analysis of a Simulated Lensed GW150914-like Event}

In this section we use \textsc{Gravelamps} to simulate a lensed waveform, with source properties modelled on those of GW150914, as an extension of the waveform generation capabilities of \textsc{bilby} -- themselves wrappers for LAL -- to give a wide variety of base, unlensed waveforms to which the generated amplification factor data can be applied. Selected here as an illustrative example is the \textsc{IMRPhenomXP} waveform model \citep{Pratten_2021}. 

To make \textsc{Gravelamps} more feature complete, it is also capable of simulating an event as one waveform, and then analysing it as another -- allowing investigation of the case where the wrong lensing profile is mistakenly applied.  Moreover, as already highlighted, \textsc{Gravelamps} is capable of generating predicted lensed waveforms under both geometric optics and wave optics scenarios.

The first analysis presented here simulates a GW150914-like event which is microlensed by a $50M_{\odot}$ mass point lens, and is correctly identified as a point-lensed event. The second analysis examines a similar scenario with a $1000M_{\odot}$ lens to show a more likely detection scenario. The final analysis explores the limitations of geometric optics alone, by showing that a $4.4 \times 10^7M_{\odot}$ Singular Isothermal Sphere, replicating a galaxy-scale `macrolensing'-type strong lensing event, analysed purely using geometric optics is unable to discriminate conclusively between SIS and NFW lens models. In all cases, the lens is placed half-way between the observer and the source, with the source position, $y$, being set to 1.  

\begin{figure*}
	\centering
	\subfloat[SIS (Correct) Model parameter estimates]{
		\label{subfig:sissimulationcorner}\includegraphics[width=0.49\textwidth]{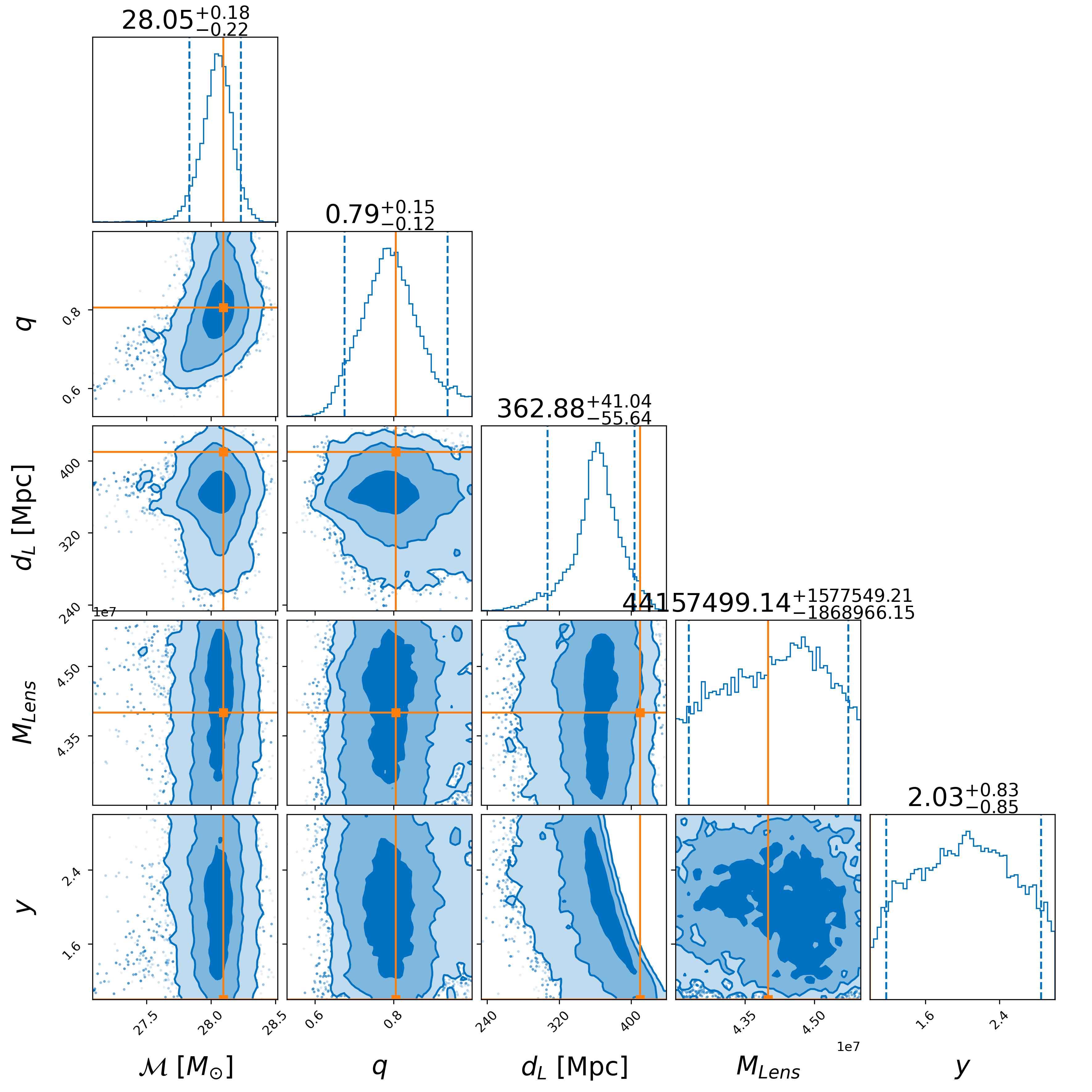}
	}
	\subfloat[NFW (Incorrect) Model parameter estimates]{
		\label{subfig:nfwsimulationcorner}\includegraphics[width=0.49\textwidth]{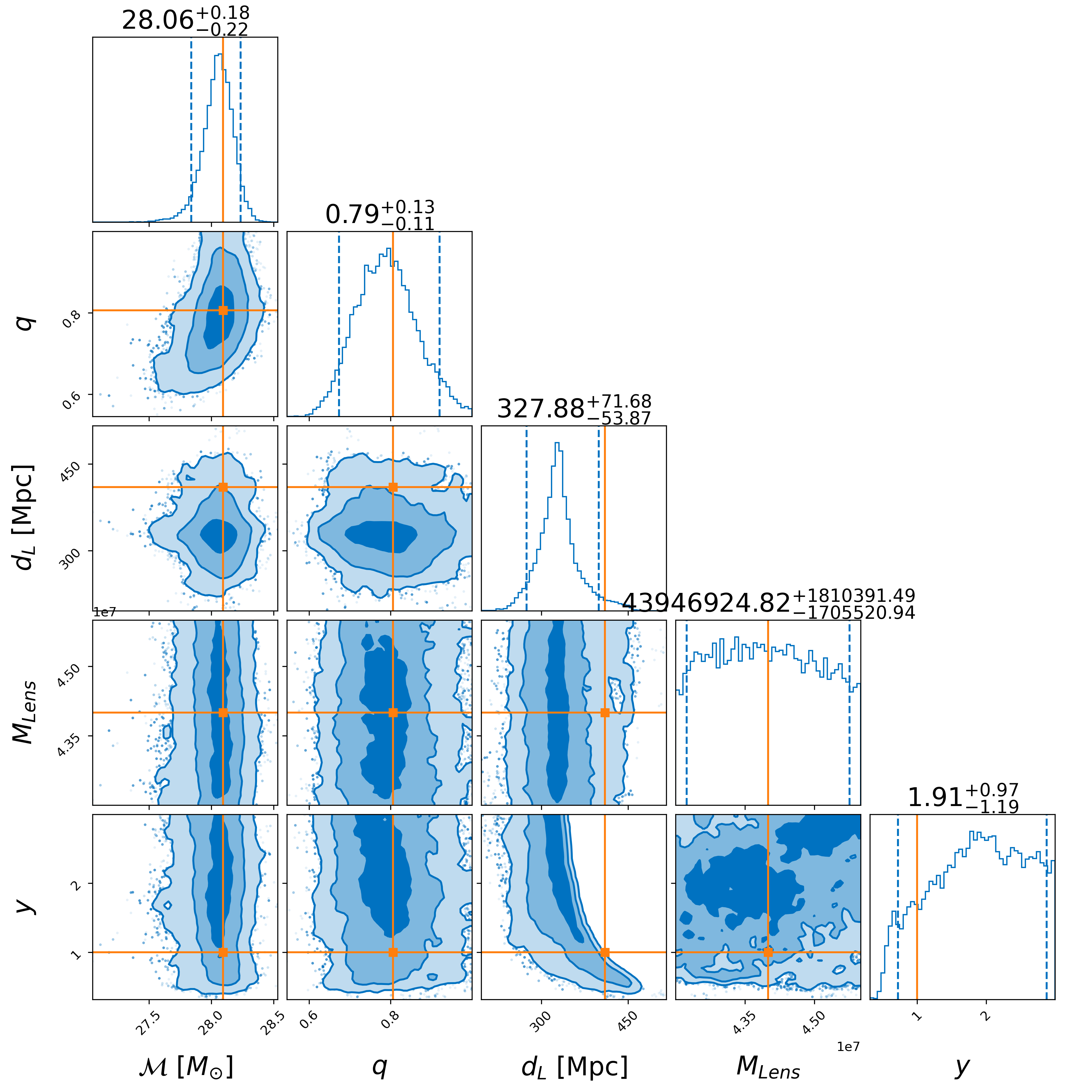}
	}
	\caption{Subset of results from lens and source model parameter estimation, performed by \textsc{bilby} nested sampling using \textsc{dynesty}, for a GW150914-like GW event macrolensed by a $4.4 \times 10^7M_{\odot}$ SIS lens, analysed as an SIS and an NFW macrolensed event respectively. As can be seen, the performance of the lens parameter estimates has decreased due to geometric optics encoding less information in the signal, however, the source parameter estimates are still constrained}
	\label{fig:macrolensing_simulation}
\end{figure*}

The true source properties used in each of these illustrative cases are summarised in Table \ref{sourpar} below. \textsc{Gravelamps} has attempted to recover all of these parameters with the exception of the geocenter time. Discussed and presented below are a subset of these fuller results considering the most obviously related parameters -- the masses, distance, and lens parameters (source frame lens mass and source position). Full configurations for the analyses presented here may be found within the Gravelamps repository \cite{gravelamps_repo} and were performed using the version located at \cite{mick_wright_2022_6371423} with the exception of those runs for Figure \ref{fig:macrolensing_simulation} which were performed with an updated version due to the identification of a bug specific to those runs during the review process.

\begin{table}
	\centering
	\begin{tabular}{c c}
		\hline 
		Parameter &  Value \\ 
		\hline
		$m_1$ & $36M_{\odot}$ \\
		$m_2$ & $29M_{\odot}$ \\
		$a_1$ & $0.4$ \\
		$a_2$ & $0.3$ \\
		$\theta_1$ & 0.5 radians \\
		$\theta_2$ & 1.0 radians \\
		$\phi_{12}$ & 1.7 radians \\
		$\phi_{jl}$ & 0.3 radians \\
		RA & 1.375 radians \\
		DEC & 1.12108 radians \\
		$d_{L}$ & 410 MPc \\
		$\psi$ & 2.659 \\
		$t_{c}$ & 1126259642.413 GPS seconds \\
		\hline
	\end{tabular}
	\caption{Source parameters of the simulated GW150914-like gravitational wave event}
	\label{sourpar}
\end{table}

\makeatletter\onecolumngrid@push\makeatother
\begin{figure*}
	\centering
	\includegraphics[width=0.8\linewidth]{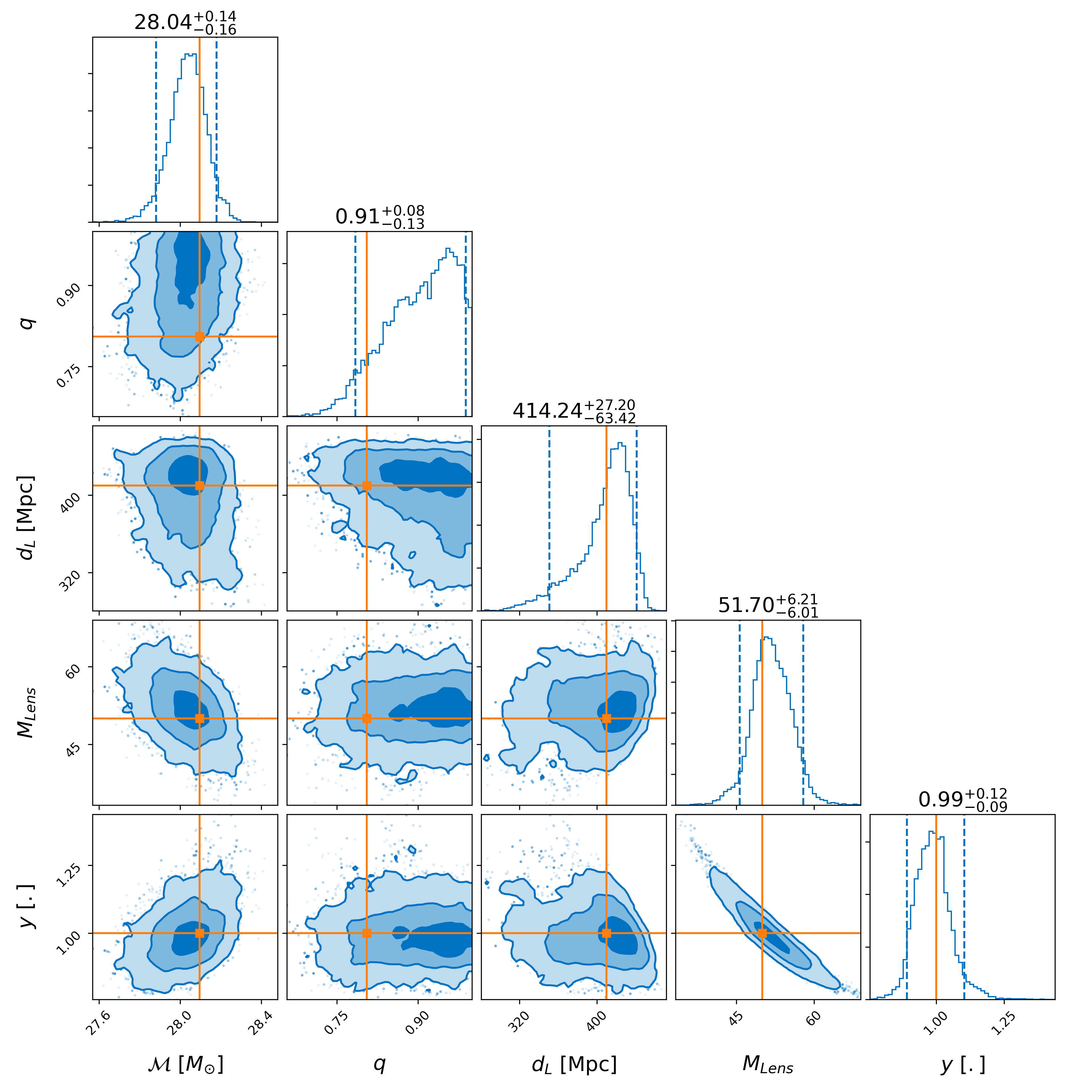}
	\caption{Subset of results of lens and source model parameter estimation, performed by \textsc{bilby} nested sampling using \textsc{dynesty}, for a GW150914-like GW event microlensed by a $50M_{\odot}$ mass point lens. Parameters shown are from left-to-right: the chirp mass ($\mathcal{M}$), the mass ratio ($q$), the luminosity distance ($d_{L}$), the source frame lens mass, and the source position. As can be seen, \textsc{Gravelamps} is able to successfully recover the model parameters.} 
	\label{microlensingcorner}
\end{figure*}
\makeatletter\onecolumngrid@pop\makeatother

Figure \ref{microlensingcorner} shows a subset of the full resulting parameter estimates inferred from the \textsc{Gravelamps} microlensing simulation when analysed with the correct model - i.e. an isolated point lens. As can be seen, each of the parameters has been recovered well, being tightly constrained to ranges consistent with their true values. This follows into the log Bayes Factor estimates - when compared with noise, analysis as a point lens, yields a value of $11063.213 \pm 0.332$. When analysis this simualated event as an SIS lensing event or a NFW lensing event with $k_s = 2$ the Bayes Factors compared with noise are $11061.194 \pm 0.334$ and $11054.524 \pm 0.345$ respectively. This leads to an overall favouring of the Point Lens by a log Bayes Factor of $2.019$ over the SIS case and $8.689$ over the the NFW case.

\makeatletter\onecolumngrid@push\makeatother
\begin{figure*}
	\centering
	\subfloat[Point Lens (Correct) Model parameter estimates]{
		\label{subfig:1000pointsimulation}\includegraphics[width=0.49\textwidth]{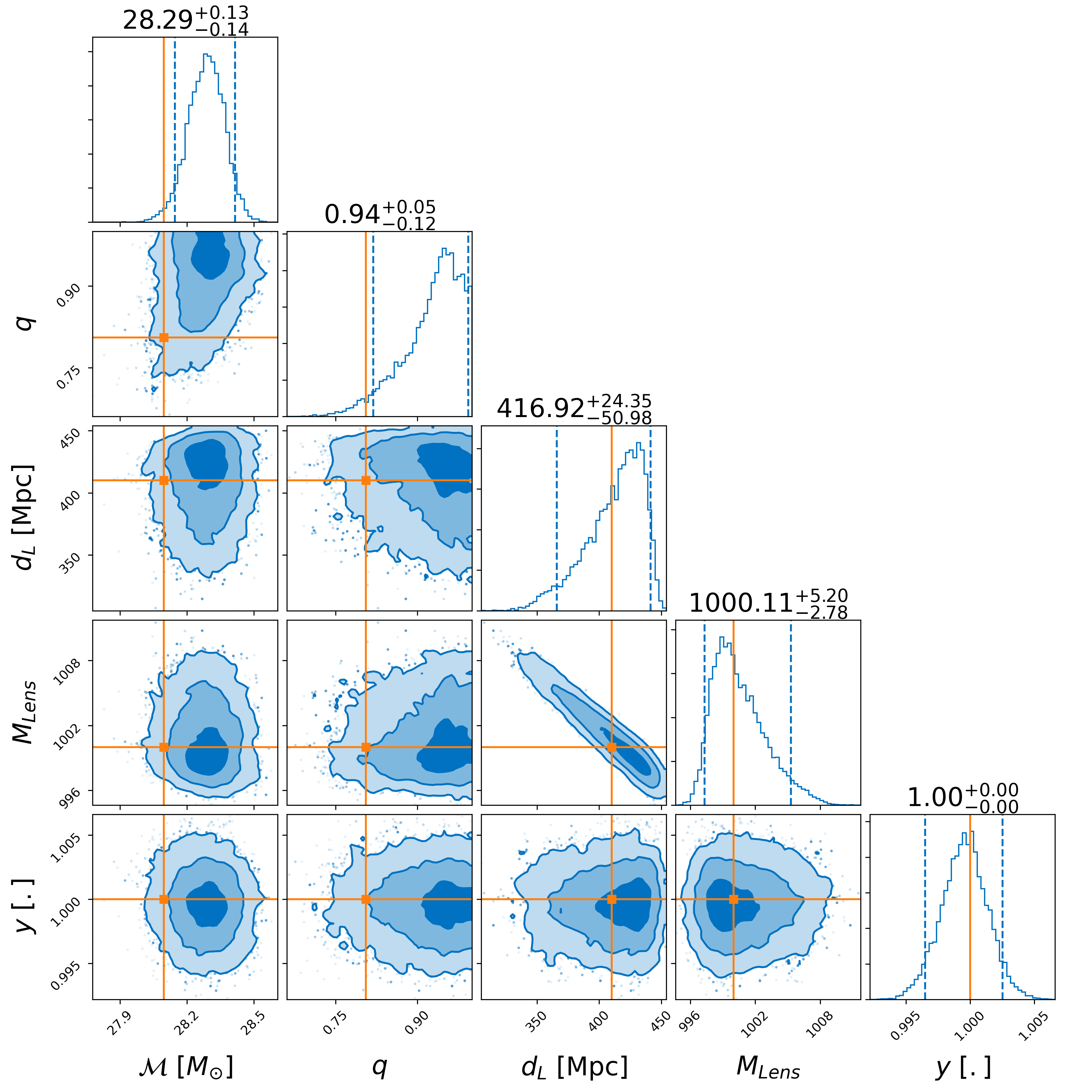} 
	}
	\subfloat[SIS (Incorrect) Model parameter estimates]{
		\label{subfig:1000sissimulation}\includegraphics[width=0.49\textwidth]{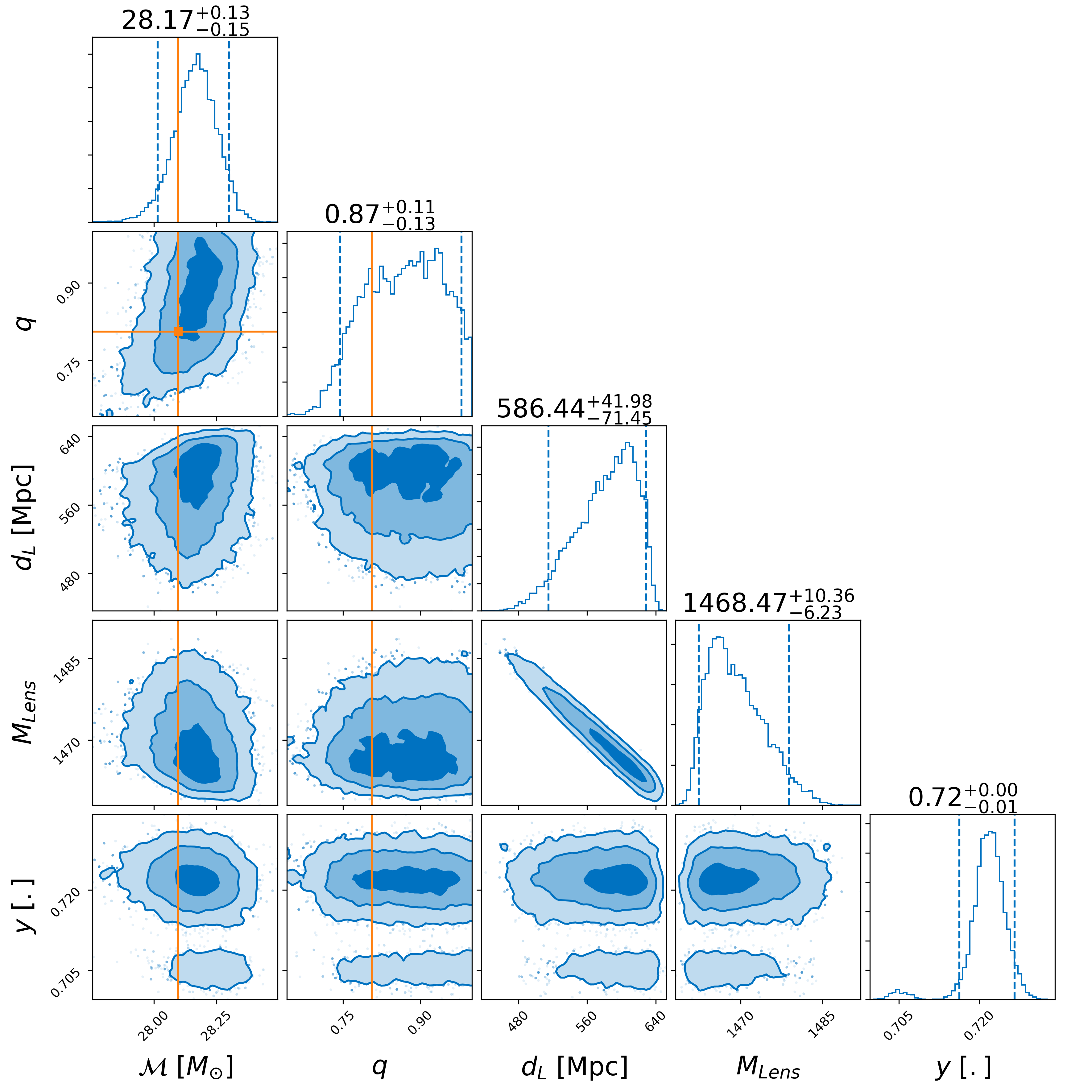}
	}
	\caption{Subset of results from lens and source model parameter estimation, performed by \textsc{bilby} nested sampling using \textsc{dynesty}, for a GW150914-like GW event macrolensed by a $1000M_{\odot}$ point lens, analysed as a point and an SIS event respectively. As can be seen, Gravelamps has again successfully recovered the model parameters in the case of the correct model.}
	\label{fig:1000_simulation}
\end{figure*}
\makeatletter\onecolumngrid@pop\makeatother

In the more likely scenario of a lensing mass on the scale of $1000M_{\odot}$ - specifically considered here was the case of $M_{\text{lens}} = 1000M_{\odot}$, the results are even more conclusive and are presented in Figure \ref{fig:1000_simulation}. In this case, the log Bayes Factor comparing the signal to the noise case was $9208.585 \pm 0.356$ for the true point lens case and $9184.879 \pm 0.352$ when analysing the simulation as an SIS event - yielding a favouring of the point lens scenario with a log Bayes Factor of $23.706$. 

\makeatletter\onecolumngrid@push\makeatother
\makeatletter\onecolumngrid@pop\makeatother

Figure \ref{subfig:sissimulationcorner} shows the resulting parameter estimates from the \textsc{Gravelamps} SIS macrolensing simulation using purely geometric optics. As can be seen, the parameter estimation performance has decreased, particualrly amongst the lensing parameters. The broadening of the lens mass posterior is particularly noticeable. This is an expected result, as the inference of ther lens mass is dependent upon the change of the amplification factor as a function of the dimensionless frequency (in which the lens mass is a contributing quantity). In the geometric optics case, the amplification factor becomes invariant over all dimensionless frequency, and the lensed signal thus becomes insensitive to the lens mass. 

Figure \ref{subfig:nfwsimulationcorner} shows the resulting parameter estimates in the case where the data is analysed under the incorrect lens model. Of particular note is that there is little difference between the parameter estimates in this case. This is reflected in the log Bayes Factors: when analysing the data with the true SIS model, the log Bayes Factor comapred with noise is $8495.794 \pm 0.241$ while for the incorrect NFW case the log Bayes Factor is $8495.065 \pm 0.242$. This yields a difference in log Bayes Factor of $0.73$ -- i.e. indicating that, when considering geometric optics alone, the two lens models are insufficiently distinguishable. 

\subsection{Analysis of Real Event GW170809}

\makeatletter\onecolumngrid@push\makeatother
\begin{figure*}
	\centering
	\subfloat[Point Mass Lens Parameter Estimates]{
		\label{subfig:pointrealcorner}\includegraphics[width=0.49\textwidth]{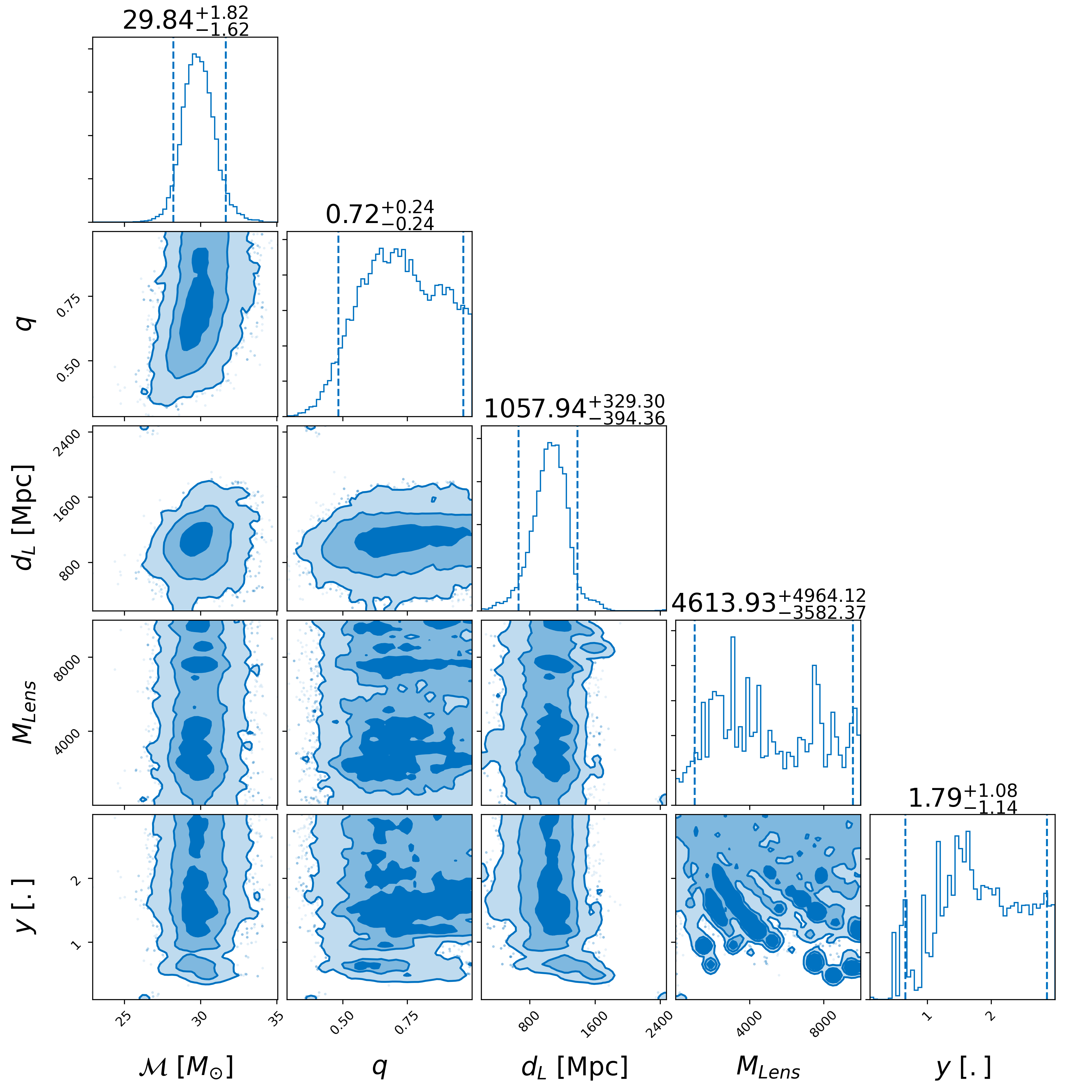} 
	}
	\subfloat[SIS Lens Parameter Estimates]{
		\label{subfig:sisrealcorner}\includegraphics[width=0.49\textwidth]{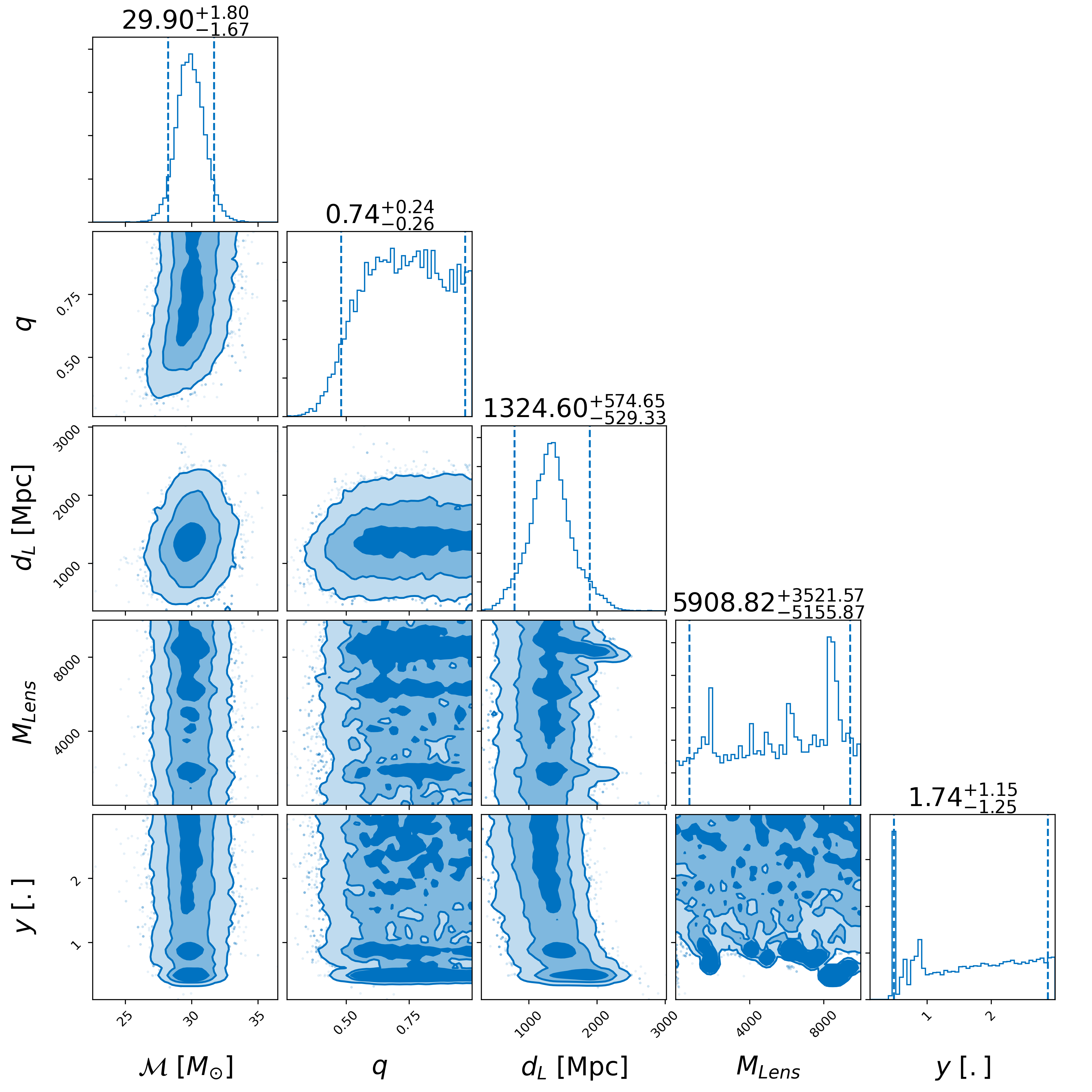}
	}
	\caption{Subset of results from lens and source model parameter estimation, performed by \textsc{bilby} nested sampling using \textsc{dynesty}, for the O2 detection GW170809. Here note the extremely broad posteriors on both the lensing mass and the source position -- which may indicate that the lensing models are not good fits to the data.}
	\label{fig:real_event_analysis}
\end{figure*}
\makeatletter\onecolumngrid@pop\makeatother

To further illustrate of the usefulness of \textsc{Gravelamps} and its suitability for analysis of real gravitational-wave event candidates, the real gravitational-wave event GW170809 was next analysed under the hypothesis that this event was lensed by a Point Mass and SIS profile respectively -- with prior ranges for each model reflecting the microlensing hypothesis.   This event was chosen due to the fact that, of those events analysed from the second observing run -- from which the data is available from the Gravitational Wave Open Science Center (GWOSC) \citep{RICHABBOTT2021100658} -- GW170809 was identified as one of the strongest (albeit ultimately rejected) lensing candidates \citep{Hannuksela_2019}. Consequently, there has also been some additional interest shown in this event as a potential lensed candidate \citep{broadhurst2019twin}. 

Figure \ref{fig:real_event_analysis} shows the result of the \textsc{Gravelamps} runs for GW170809 analysing in the microlensing regime for the point (\ref{subfig:pointrealcorner}) and SIS (\ref{subfig:sisrealcorner}) models respectively. In this case, the log Bayes Factors -- comparing the hypothesis of modelled signal + noise, versus that of pure noise -- for the point and SIS cases are $56.773 \pm 0.265$ and $57.138 \pm 0.258$ respectively.  We see, therefore, that the signal + noise hypothesis is strongly favoured -- consistent, of course, with the fact that GW170809 was identified as a confident gravitational-wave detection on GWTC-1. In this case, however, the lensing parameter estimates are particularly broad as compared with the simulations which may be an indication that the lensing models are ill-fitting to the data, an uninformative posterior in the lensing parameters being what one would expect in the case of an unlensed signal.  Thus, while our \textsc{Gravelamps} analysis does marginally favour the SIS lens model over the point lens model with a log Bayes Factor of $0.365$, these results should be taken as merely illustrative of the capabilities of the software rather than any indication of a preference for a particular lens model for GW170809. We emphasise again that \textsc{Gravelamps} does not explicitly take into account the selection effects, multiple image analysis, and more detailed population prior choices that are explored in \citep{theligoscientificcollaboration2021search}, in order to better assess whether there is evidence supporting the lensing hypothesis for any given gravitational-wave candidate event. Nevertheless, we believe that the example of GW170809, and the previous simulated examples, illustrate the efficacy of \textsc{Gravelamps} for comparing the evidence for different lens models \textit{given} that a lensed gravitational-wave event has been detected.

\section{Conclusion}
With the increasing sensitivity of the existing ground-based detectors, and the additional detectors set to join the global network in the future, the detection of a lensed gravitational-wave event within a matter of a few years would appear to distinctly possible. Those searches performed thus far have focused on identifying candidate lensed events; however, once such a candidate event has found, the immediately important question becomes what is the astrophysical nature of the lens itself?  The package \textsc{Gravelamps} presented here has been designed to help answer that question. 

\textsc{Gravelamps} has been designed to be an easy to use and versatile platform, with the flexibility to allow investigation of both microlensing scenarios and the so-called intermediate `mililensing` case. \textsc{Gravelamps} is particularly adapted to this latter case by having the flexibility to calculate the lensing amplification factors in either geometric or wave optics, or in a hybrid combination of both, over the course of a single analysis run -- with full control given to the user as to where each regime may apply. As an initial, illustrative, set of lensing models, \textsc{Gravelamps} presently contains the point mass, Singular Isothermal Sphere, and Navarro, Frenk, White density profile; however it has been designed to easily be extended to include any lens model that the user wishes to investigate. 

We have demonstrated the utility of \textsc{Gravelamps} by presenting examples, in both of the wave and geometric optics cases of the results of parameter estimation. Here we demonstrated that one of the major limits of the geometric optics case is that alone it is insufficient to determine the lensing model - at least within the single image analysis performed here. However, scenarios which may partially incorporate geometric optics cases alongside wave optics effects - looking for lensing in the $\mathcal{O}(1000M_{\odot})$ region for instance - are identifiable. 

As noted, \textsc{Gravelamps} is an easily extensible platform and the work presented here is seen as the just the first step in its development. For example, in future we plan to explore its extension to handle multiply lensed signals -- for instance the case of a source that simultaneously undergoes both microlensing and macrolensing. 

\begin{acknowledgments}
	\section*{Acknowledgements}
	The authors acknowledge computational resources provided by Cardiff University, and funding provided by STFC to support UK Involvement in the Operation of Advanced LIGO. MW acknowledges funding support provided by the Stirlingshire Educational Trust and the Scottish International Education Trust. MW is also grateful for the generous financial contribution of Mrs Catherine Goldie to this research.  MH acknowledges additional support from the Science and Technology Facilities Council (Ref. ST/L000946/1). This research has made use of data, software and/or web tools obtained from the Gravitational Wave Open Science Center (https://www.gw-openscience.org/ ), a service of LIGO Laboratory, the LIGO Scientific Collaboration and the Virgo Collaboration. LIGO Laboratory and Advanced LIGO are funded by the United States National Science Foundation (NSF) as well as the Science and Technology Facilities Council (STFC) of the United Kingdom, the Max-Planck-Society (MPS), and the State of Niedersachsen/Germany for support of the construction of Advanced LIGO and construction and operation of the GEO600 detector. Additional support for Advanced LIGO was provided by the Australian Research Council. Virgo is funded, through the European Gravitational Observatory (EGO), by the French Centre National de Recherche Scientifique (CNRS), the Italian Istituto Nazionale di Fisica Nucleare (INFN) and the Dutch Nikhef, with contributions by institutions from Belgium, Germany, Greece, Hungary, Ireland, Japan, Monaco, Poland, Portugal, Spain. This material is based upon work supported by NSF’s LIGO Laboratory which is a major facility fully funded by the National Science Foundation. 
\end{acknowledgments}
\nocite{*}
\bibliographystyle{aasjournal}
\bibliography{ShortAuthRefs}
\end{document}